# BRAIN REPRESENTATIONS OF PERCEPTUAL STIMULI AT DIFFERENT LEVELS OF AWARENESS

**Birgitta Dresp**

ICube UMR 7357 CNRS-UdS
Strasbourg-FRANCE

**Abstract**

This article questions the widespread assumption that there are brain representations that will always remain unconscious in the sense of being « inaccessible to an individual's awareness under any circumstances ». This implies that some part of the knowledge generated by the brain is once and for always excluded from an individual's consciousness and, therefore, from being communicated to the outside world. This standpoint neglects the possibility that the human brain might have a capacity for generating meta-representations of non-conscious knowledge contents at a given moment in time through context-sensitive adaptive learning, and is somewhat difficult to reconcile with experimental findings showing that initially subliminal targets can be made available to awareness, or « break through » to supraliminal levels of processing, when they are embedded in the appropriate perceptual object context. Functional properties of neural network architectures inspired by the functional organization of the primate cortex are able to explain how a human brain could generate this kind of perceptual learning. Signals or knowledge processed outside awareness can be made available to awareness through adaptive resonance of bottom-up and top-down signal exchanges in massively parallel neural network architectures, in other words, on the basis of statistically significant signal matches in the domain of time and in the domain of contents.

# Introduction

The majority of scientific approaches in the behavioral and brain sciences investigates mechanisms and processes that are activated by supraliminal experimental stimuli. As a consequence, the expression of these mechanisms and processes in the behavioral and neural data reflects a level of integration well within the assumed capacity limits of the system under investigation. A major problem arising from such a « supraliminal approach » to perceptual and cognitive function is that our understanding of a given system can be neither complete nor accurate without knowledge about its subliminal capacities, or its processing limits. Only once we are able to assess the full range of processing capacities of a brain system such as a sensory system, recognition system, or a memory system, we can define and predict the characteristics of the input necessary to get the system to operate efficiently, and only then can we make reliable assumptions about the nature of the mechanism(s) we are supposed to be investigating.

Scientific evidence for the existence of subliminal perceptual processes, which means here mechanisms and processes activated by stimuli a human observer is not aware of, has accumulated over the last three decades. Some of these data, especially the most recent, will be reviewed in the first chapter of this article. They converge in establishing that subliminal perceptual and representational processes which operate outside awareness can be made available to perceptions and representations embedded in immediate, ongoing awareness at various levels of processing. Most of them can be interpreted as evidence for parallel distributed sensorial and representational processes that operate within separate streams at different levels of awareness with interactions at the sensorial, representational, and memory level. Altogether, current and earlier research relating to subliminal perceptual processes raises issues regarding their functional significance, whether processes occuring outside awareness should be regarded as qualitatively different from processes within the domain of awareness, and whether there are brain mechanisms that may allow to account for shifts in levels of awareness. These questions are discussed in the second chapter. A working hypothesis, or theory, for subliminal perceptual and representational processes is elaborated on the basis of current neural network theory (Carpenter & Grossberg, 1991 ; Grossberg, 1999) and confronted with earlier theories of conscious and unconscious cognitive worlds (Kihlstrom, 1987). The working hypothesis states that subliminal and supraliminal signals and representations are processed in parallel, and that subliminal representations can be made available to awareness via temproal coincidence within an appropriate context via mechanisms that bind distributed network information at multiple levels of brain processing into context-sensitive representations of knowledge and events. It will be assumed that the emergence of an aware representation of subliminal knowledge involves interactions between neural structures that receive actual bottom-up input, and neural structures that do not receive bottom-up input. These latter can become activated when subliminal representational traces match the input from connected structures that receive, process, and transmit current bottom-up signals which may themselves be subliminal.

By « subliminal » we commonly have to understand « not consciously perceived ». In some cases, however, a technically more precise definition is given. In visual or auditory processing, for example, a subliminal stimulus would be defined as one « below the detection threshold ». The detection threshold (e.g. Green & Swets, 1966) is arbitrarily defined as the stimulus intensity that is needed to yield 75% correct judgements about the presence or absence of the stimulus in a given number of trials in which it is presented. Such a definition appears to make sense as detection thresholds usually vary only little, for one and the same as well as between observers, it does, however, make no statement about whether or not conscious perception of the subliminal stimulus is possible. In general, it can be assumed that a stimulus that is not detected in 75 % of a given number of trials would most of the time not be consciously perceived either (e.g. Merikle & Reingold, 1990).

A great deal of research on subliminal perceptual processes uses the less technical definition of the concept, and this may represent a problem regarding the generality of the findings. However, the existing body of data is interesting and challenging enough to deserve a detailed overview and a discussion. The evidence for subliminal knowledge representations and their possible interaction with supraliminal processes concerns various levels of affective, cognitive, and sensory processes. This is revealed by memory, learning, or sensory performances observed in the absence of any recalled experience or phenomenal awareness as in subliminal semantic priming, for example, and by experiments on interactions between subliminal and supraliminal visual processing.

**Subliminal perceptual processes : Behavioral and neurophysiological evidence**

Experimental, clinical, and neurophysiological data which bring to the fore a crucial role of subliminal perceptual processes in affective, cognitive, and sensory function are reviewed in the following paragraphs. In regard to subliminal sensory coding, vision only will be considered here for lack of data on other sensory modalities. The phenomena that will be reviewed and discussed here are subliminal psychodynamic activation (e.g. Silverman, 1983), shifts from supraliminal to subliminal perception in hypnosis (e.g. Chaves & Dworkin, 1997), subliminal semantic priming (e.g. Marcel, 1989) and the effects of undetected context stimuli on recognition processes, subliminal learning (e.g. Wong, Bernat, Bunce, & Shevrin, 1997), the processing of subliminal visual signals as a function of immediate visual context (e.g. Dresp, 1998), and phenomena of blindsight in patients with striate cortical lesions (Weiskrantz, 1986), in non-human primates (Cowey & Stoerig, 1995), and in normal observers (Kolb & Braun, 1995).

*Subliminal psychodynamic activation*

Subliminal psychodynamic activation generally describes behavioral effects where the exposure to subliminally presented, drive-related stimuli results in a positive change in the emotional and mental state of human observers (Silverman, 1983; Silverman & Weinberger, 1985). In particular, results from clinical studies have shown that subliminal verbal messages designed to induce symbiotic fantasies and administered under double-blind quasi-experimental conditions significantly reduce anxiety levels and raise the motivation of psychiatric patients such as drug abusers (Thornton, Igleheart, & Silverman, 1987). Follow-up examinations furthermore revealed that the experimental patient groups who received treatment with the subliminal stimuli reported more dreams containing positive symbiotic events than the controls.

It is emphasized that the non-conscious character of the stimuli in subliminal psychodynamic activation (SPA) is critical. Effects produced under conditions where observers are unaware of the nature and content of the stimuli were found to be significantly stronger than those produced by the same stimuli presented at supraliminal levels (Bornstein, 1990). Explanatory models of SPA effects suggest that supraliminal stimuli lose some of their power to produce the desired effects on internal representations because subjects perceive them as part of an externally administered procedure (Bornstein, 1992). In other words, stimulus awareness would in this case diminish the organisms capacity for responding efficiently to drive- and affect-related stimuli. Some restricting effect of awareness on psychodynamic responsiveness is widely believed to diminish the efficiency of relaxation techniques that combine soothing music with verbal suggestions, which has lead to the sustained use of subliminal suggestions combined with soft music in relaxation therapy.

Experimental studies (e.g. Chaloult, Borgeat, & Elie, 1988) have shown that the most efficient combinations appear to be indeed those where soft music is presented together with verbal stimuli of an intensity slightly below the level of conscious perception.

Theory and findings regarding SPA effects have received critical feed-back raising issues relating to the appropriateness of control and threshold stimuli in the various experiments (Malik, 1998 ; Malik, Krasney, Aldworth, & Ladd, 1996), the possible need for physiological indicators of anxiety reduction such as the subject's heart rate in addition to the psychological measures (Malik, Paraherakis, Joseph, & Ladd, 1996), and questions about the need for neutral, i.e. neither drive- nor affect-related, stimuli to establish individual subjective thresholds for SPA (Greenberg, 1998 ; Malik, 1998). However, quantitative and qualitative reviews and meta-analyses of subliminal symbiotic activation research that has been conducted over the years led to the conclusion that, despite some possible artefacts, the main results remained statistically significant (Hardaway, 1990), partial-cue hypotheses of SPA effects were not tenable (Bornstein, 1990), and that the findings preserve their full implication for cognitive science, subliminal perception research, and any research that is to examine the negative influence of awareness on responding optimally to drive- and affect-related stimuli (Bornstein, 1990; Hardaway, 1990;  Malik & Paraherakis, 1998).

In relation to subliminal psychodynamic activation phenomena, hypnosis and hypnotic suggestibility also deserve some attention. Although hypnosis, or more precisely hypnotic induction, are not subliminal phenomena *per se,* as the induction of psychodynamic effects in hypnosis is mediated via esentially supraliminal verbal suggestions, it is agreed that hypnotic phenomena may be best described and understood as products of an altered state of awareness (Chaves, 1997). If this is conceivable, and given what is assumed about the possibly restricting effects of awareness on psychodynamic responsiveness, hypnosis research might have some relevance in regard to subliminal perceptual phenomena.

The degree to which a human individual may respond to hypnotic suggestions is referred to as hypnotic susceptibility and can be accurately predicted on the basis of psychometric tests such as the Waterloo-Stanford Group Scale of Hypnotic Susceptibility (Bowers, 1993 ; 1998). Hypnotic suceptibility is an estimate of the ability of a man or a woman to enter some trance-like state where overall awareness is shifted away from the general context and environment, and focussed on the symbiotic fantasies induced by the hypnotic (verbal) suggestions of an expert clinician. Hypnotic suggestibility in young men and women has been shown to be significantly enhanced following application of weak (1 micro Tesla) burst-firing magnetic fields for 20 minutes over the right temporoparietal lobes

(Healey, Persinger, & Koren, 1996). The findings suggest that the signatures of these low-frequency magnetic fields contain biorelevant information which directly affects the neural processes underlying hypnotizability. Positron emission tomography (PET) measures of regional cerebral blood flow and electroencephalographic (EEG) measures of brain electrical acitvity have shown that specific patterns of cerebral activation are associated with the hypnotic state and the processing of hypnotic suggestions (Rainville, Hofbauer, Paus, Duncan, Bushnell, & Price, 1999). Another PET study comparing highly susceptible males with an additional ability to hallucinate under hypnosis, so-called hallucinators, to other highly hypnotizable non-hallucinators revealed that a specific region in Brodman area 32 was activated in the group of hallucinators when they heard an auditory stimulus or when they merely hallucinated hearing it under hypnosis (Szechtman, Woody, Bowers, & Nahmias, 1998). Such an activation was absent when the hallucinators simply imgined hearing the tone, and in all experimental conditions with the group of non-hallucinators.

Measurable consequences of hypnosis intervention on cognitive function were reported. With highly susceptible observers, hypnosis produces an inhibition of correct responses in perceptual tasks with conflicting stimulus information (Kaiser, Barker, Haenschel, Baldeweg, & Gruzelier, 1997) correlating with changes in error-specific negativity and positivity of cortical evoked potentials. Effects of hypnotic susceptibility on auditory event-related potentials (AERPs) were found with observers who were instructed to ignore tones while accomplishing some other task such as reading a novel. The highly hypnotizable subjects revealed different AERP amplitudes and latencies when ignoring the tones, and were significantly slower in responding to the not-to-be-attended stimuli than the less susceptible subjects. The findings are interpreted as evidence that highly hypnotizable humans have a greater ability to shift awareness towards relevant stimuli and away from irrelevant ones (Crawford, Corby, & Kopell, 1996). Furthermore, specific hypnosis techniques such as suggested selective deafness or selective visualization appear to influence learning processes in the desired direction (Rodriguez-Sanchez et al., 1997).

A particular example showing how supraliminal perceptions or representations may sometimes become genuinely subliminal through guided shifts in awareness induced by hypnotic suggestions is the hypnotic control of physical pain, referred to as hypnotic analgesia (Hilgard & Hilgard, 1983 ; Chapman & Nakamura, 1998). Overall interest in and scientific information about hypnotic analgesia appears to have grown substantially in recent years, which has had significant influence on strategies for acute and chronic pain management in the private and public domain. Although it is often difficult to distinguish facts from artefacts such as placebo and similar phenomena, the state-of-the-arts in the domain points towards some general agreement that pain and distress perception in acute as well as chronic pain patients with high hypnotic suceptibilty can be significantly lowered through hypnosis (see Chaves & Dworkin, 1997, for a review). Recent scientific studies have investigated the effect of hypnotically induced obstructive fantasies to somatosensory stimuli (hypnotic analgesia) on pain and distress tolerance ratings, EEG spectral amplitude, heart rate, and P300 event-related potential amplitudes. The results of these studies showed significantly better pain and distress tolerance, significant changes in EEG amplitude, and a significantly reduced heart rate (De Pascalis, & Perrone, 1996) in highly susceptible subjects following painful electrical stimulation under hypnosis. P-300 amplitude peaks to standard somatosensory stimuli was found to be significantly reduced in subjects with high hypnotizability in a pain-target detection task (De Pascalis & Carboni, 1997).

Apart from the generally growing refinement in hypnosis research and its possible implications for clinical issues, hypnotic phenomena also have an undeniable, though not yet fully appreciated, significance within a cognitive neuroscience perspective as they provide evidence that perceptions and representations embedded in immediate, ongoing awareness are fed into parallel distributed processes that operate outside awareness (Chapman & Nakamura, 1998 ; Kihlstrom, 1998).

*Subliminal semantic priming and associative learning*

The question whether a person 's feelings, judgements, or choices can be influenced by subliminal images or messages while watching television or looking at an advertisement (Simpson, Bown, Hoverstad, & Widing, 1997) has been subject to discussion for quite some time. In a BBC broadcast study presented in 1994, faces were flashed subliminally within the programme for about 20 milliseconds in a restricted part of the network region. Immediately after the broadcast, TV viewers were invited to make a judgement by telephone about a neutral, supraliminal face image that expressed no emotion. Judgements were made by telephoning one of two numbers (1 or 2) indicating « sadness » or « happiness ». Statistical analyses of the phone call responses revealed that viewers who received a subliminal smiling face in the broadcast were less likely to judge the neutral face as being happy than were those viewers who were not exposed to the subliminal image in the programme. Underwood (1994) suggested that this effect could be explained in terms of some kind of contrast effect where the neutral expression of the supraliminal image is interpreted as « sadder » than the smiling subliminal image. However, the broadcast study provided no information as to whether the so-called subliminal frames could have been perceptible in some cases, i.e. available to awareness, and attempts to replicate the results of the broadcast study under laboratory conditions (Underwood, 1994) did not yield findings unambiguous enough to allow for a clear conclusion.

Scientific evidence for truly subliminal perceptual processes in recognition, memory, and learning phenomena dates back to work by Marcel (1983) using experiments investigating the effects of visual masking on word recognition. These earlier findings, suggesting that supraliminal perception is not necessary for recognition, are confirmed by results from several more recent studies using subliminal priming paradigms. Subliminal priming describes an experimental technique where target stimuli are preceded by non-perceptible stimuli, so-called primes, which are supposed to have a deterministic influence on the processing, or recognition, of the supraliminal targets. Experiments using near-threshold visual primes in a memory task with brief flashes of previously non-recalled items as prime stimuli have shown that near-threshold primes significantly increased the number of recalls of otherwise non-recallable items although the « feeling of knowing » reported by the observers did not change (Jameson, Narens, Goldfarb, & Nelson, 1990). These findings provide further evidence that subliminal perception efficiently modifies representations in ongoing awareness while the process itself as well as its immediate behavioral outcome may remain outside awareness. More evidence for subliminal semantic priming comes from experiments where subjects had to classify visually presented words (targets) into semantic categories. Prime words were rendered more or less subliminal through masking and brief exposure durations between 17 and 50 milliseconds, and observers were instructed to respond within a narrow time window (Draine & Greenwald, 1998). The magnitude of priming effects as a function of prime detectability (Greenwald, Klinger, & Schuh, 1995) was assessed using linear regression analysis. Substantial effects of semantically congruent but truly subliminal primes reflected by significantly lower error rates were reported (Draine & Greenwald, 1998). The results from these experiments converge in establishing that semantic processing is effective at subliminal levels, as indicated by earlier priming studies producing evidence for memory and recognition beyond awareness (e.g. Marcel, 1983 ; Greenwald, Klinger, & Liu, 1989).

A recent neurophysiological study has provided evidence for brain correlates of semantic priming, using a combination of behavioral task and brain-imaging technique (Dehaene, S., Naccache, L., Le Clec, Koechlin, Mueller, Dehaene-Lambertz, van de Moortele, & Le Bihan, 1998). It was shown that subliminal prime stimuli have a measurable influence on electrical and haemodynamic correlates of brain activity. Other functional neuro-imaging studies have investigated brain correlates of the « mere exposure effect » (Shevrin, Smith, & Fritzler, 1971; Kunst-Wilson & Zajonc, 1980 ; Bornstein & D'Agostino, 1992), which describes the observation that mere pre-exposure to subliminal visual stimuli, non-perceptible in the sense that recognition is performed at the chance level, is sufficient to significantly influence subsequent preference and memory judgements. The subliminally triggered « mere exposure effect » may be seen as a variation of subliminal semantic priming since it reflects, like priming effects in word recognition, some direct influence on memory judgements. Neural activity in right lateral prefrontal cortex associated with implicit memory retrieval (Elliot & Dolan, 1998) was found in groups of subjects making memory and preference judgements regarding supraliminally presented objects after previous exposure to subliminal stimuli. The data are interpreted in terms of a significant activation of a memory system in the absence of recollective experience, i.e. awareness that the preferred or memorized stimuli have been seen before. These findings appear to be consistent with earlier evidence for right lateral prefrontal activation during implicit behavioral guidance without awareness (Berns, Cohen, & Mintun, 1997).

To investigate whether associative learning occuring outside awareness could be tagged by a specific brain activity, event-related potentials (ERP) in classical aversive conditioning to subliminally presented faces via behavioral electroshock-versus-no-shock techniques were compared to ERP activity with aversive conditioning to supraliminally presented faces (Wong, Bernat, Bunce, & Shevrin, 1997). ERP activities indexing the acquisition of a conditional response to the subthreshold stimuli were found, indicating that brain traces of classical conditioning are established in the absence of awareness. Behavioral evidence for subliminal learning has been made available in experiments showing that the subliminal presentation of one of two contingent stimuli in a choice reaction time task is sufficient, with some training, to yield the same reaction time pattern as the presentation of two supraliminal signals (Wolff & Rübeling, 1994). It was made sure that the observers were not aware of the stimulus contingency during the training phase, i.e. unable after training to report the contingency to which they were exposed to. Reinforcing effects of subliminal auditory affirmations embedded in soft background music on learning 'face-name-occupation' lists have been reported (Chakalis & Lowe, 1992). Subjects who were exposed to the additional subliminal input did significantly better than controls in recalling names from the list. Other learning phenomena, where the affective evaluation of a previously neutral stimulus is changed after association with a second, positive or negative, affective stimulus, have been reported with subliminal stimuli, suggesting that the evaluative associations can be learnt without knowing (De Houwer, Hendrickx, & Baeyens, 1997). In general, scientific evidence for subliminal perceptual and representational processes provided on the basis of masked priming and subliminal learning effects thus far seems to suggest that neither the effective recognition and memorization of external events, nor the measurable traces of these processes in brain activity, do necessarily require awareness.

*Subliminal vision*

Discussing the role of subliminal processes at the sensory level requires making a clear distinction between the sensory threshold, that is the psychophysical, or statistical, threshold for the detection of a stimulus as defined by Signal Detection Theory (Green & Swets, 1966), and further thresholds implied in the semantic processing of a stimulus, such as recognition thresholds. The visual sensory modality only will be considered here in this article. The conceptual distinction between detection and recognition is of crucial importance, namely in view of considerations regarding awareness and what may be assumed about shifts in awareness possibly underlying effects where previously subliminal visual stimuli suddenly become supraliminal within a specific context. A subliminal visual stimulus is defined here as a stimulus that is presented at intensity levels below the psychophysical detection threshold. During exposure to a psychophysically subliminal stimulus in a visual task, a human observer may sometimes be aware of the fact that he/she has seen something but will not be able to say what it was (Dresp, 1990), or remain unaware of the specific characteristics of the stimulus. Exposure to psychophysically supraliminal stimuli may yield recognition of what was seen in some of the trials, but not necessarily in all of them. How often recognition of supraliminal stimuli occurs will depend on the supraliminal intensity chosen in the experiment. It has been some time that the influence of subliminal stimuli on mechanisms of spatial and temporal integration of contrast has been investigated psychophysically (Battersby & Defabaugh, 1969 ; Herrick, 1973 a, b ; Kulikowski & King-Smith, 1973), showing that subliminal input matters in vision. Recent electrophysiological studies have shown significant event-related brain responses to subliminal visual stimuli (Brazdil, Rektor, Dufek, Jurak, & Daniel, 1998), and that a specific component of the P-300 brain wave could be assigned to the processing of a subliminal visual target. Evidence for a shift from subliminal to supraliminal processing as a function of visual context indicating that, at the level of sensory processing, shifts in awareness may be triggered by changes in the nature of the visual input, has been provided in psychophysical experiments.

A visual stimulus that remains undetected (subliminal stimulus) when it is presented to human observers on a blank screen may become indeed detectable (supraliminal) when it is embedded within the appropriate visual context. Detection thresholds for a small target light spot have been shown to decrease considerably when the target is presented collinear to a visual context such as a thin line or an edge (Dresp, 1993), or when it is presented collinear to the edges of visual configurations that give rise to the perception of so-called illusory contours (Dresp & Bonnet, 1995). Detection facilitation effects engendered by visual contexts have also been observed with line stimuli as targets (Kapadia, Ito, Gilbert, & Westheimer, 1995 ; Dresp & Grossberg, 1997). In most of these studies, the targets remained undetected, as reflected by detection performances below threshold, when presented without the context stimuli. Subliminal colour line targets were found to become supraliminal when presented in an appropriate colored context, but needed slightly longer exposure durations for the effect to occur than achromatic versions of the same stimuli (Dresp & Grossberg, 1999; Dresp, 1999).

Neural correlates for facilitated visual detection through collinear context structures have been found in V1 of an awake behaving monkey accomplishing the psychophysical detection task (Kapadia, Ito, Gilbert, & Westheimer, 1995). Neural activity triggered by the target alone was found to be increased further by the presence of the facilitatory visual context, and diminished by the presence of a non-facilitatory context. On the basis of these correlates and further neurophysiological evidence for effects of neural activation or suppression « beyond the classic receptive field » (Gilbert & Wiesel, 1990 ; Gilbert, 1998), it may be assumed that interactions between neural mechanisms involved in the processing of a subliminal target and neural mechanisms involved in the processing of a supraliminal context underly changes in target detectability produced by the context.

Shifts in visual detectability from subliminal to supraliminal levels reflect changes in awareness of the characteristics of the target stimulus. As long as the target remains subliminal, observers are unable to report whether what they may have seen is dark or light in the case of achromatic stimuli (Dresp, 1990), or to tell the colour of a chromatic target. When the target becomes supraliminal via the added context, this information becomes available because the observer then is able to say that he/she has seen a red target, for example (Dresp & Grossberg, 1999). Furthermore, increased target detectability produced by a visual context was found to be highly sensitive to practice effects, in other words the target was detected increasingly better with the progression of the trial blocks whereas in the condition without context, it was found to remain subliminal (Dresp, 1998). Thus, investigating effects of visual context on the processing of subliminal target stimuli brings to the fore different levels of awareness in visual processing that may be mediated via differential amounts of neural interaction activated by the visual input itself and possibly other input that is directly relevant to the visual task.

Evidence that other pathways than those projecting to striate cortex are involved in vision, and that a great deal of visual processing takes place outside awareness, comes from studies on so-called blindsight phenomena. The classic blindsight phenomenon describes behavior in patients with cortical blindness caused by lesions to their primary visual cortex (striate cortex V1), revealing residual visual capacity in the absence of the ability to report what they perceive (Weiskrantz, 1986). Research on blindsight has shown that these patients accurately detect monochromatic visual stimuli and patterns, can discriminate direction of movement as well as orientation of stimuli in their « blind » field, and are able to discriminate the wavelength of chromatic stimuli in the absence of any acknowledged perception of colour, the phenomenal attribute of chromatic stimuli (Weiskrantz, 1996, for a review). Whether the loss of phenomenal vision, or perceptual awareness, is a necessary consequence of striate cortical destruction has not yet been clarified (Stoerig & Cowey, 1997). In recent experiments, two patients with homonymous right hemianopias were tested in a number of perceptual tasks designed to assess shape perception in objects presented within the blind field (Marcel, 1998). The results show that the two observers were capable of making appropriate preparatory manual adjustments to grasp objects presented in the blind field, were consistently semantically biased (semantic priming) to a significant degree by words presented in the blind field, and were able to process structural stimulus components and the spatial order of letters presented in the blind field. Both observers were furthermore able to report afterimages of figures presented in the blind and sighted fields provided the two images together formed a Gestalt. The conclusion drawn from this study was that blindsight does not affect the use of shape percepts in motor control, that the main deficit in blindsight appears to be one of awareness not one of visual function, and that the loss of visual awareness in the blind field is not total (Marcel, 1998). The blindsight phenomenon has been found in monkeys with unilateral removal of V1 (Cowey & Stoerig, 1995) demonstrating residual visual capacity in the sense that the animals still can detect and localize visual input in their affected hemifields, but do not seem to be able to identify the nature of this input.

In normal observers, an observation similar to the blindsight phenomenon was reported in experiments using a localization task with target stimuli supposed to activate V1 but not area MT (Kolb & Braun, 1995). It was shown that the subjects successfully performed in the localization task although they were not aware of having seen the target. Furthermore, there was no correlation between success in the localization performance and subjective confidence ratings. With target stimuli available to awareness, a positive correlation between these two variables was found. Altogether, the data on blindsight phenomena in human and non-human primates reveal that efficient visual processing is possible in the absence of awareness, and that humans are capable of making appropriate preparatory motor adjustments to visual objects they are not aware of. One question that arises from the evidence for subliminal perceptual processes is « what are these processes there for » ? The following chapter will provide an attempt to answer this question, and introduce the core hypotheses of this article.

**Does the brain know more than we do ? Towards a working hypothesis for subliminal perceptual processes**

Data on altered states of awareness, on the restricting effects of awareness on psychodynamic responsiveness, and on mental states where overall awareness is shifted away from the immediate environment, or from painful sensations, and focussed on the symbiotic fantasies induced by the hypnotic suggestions of an expert clinician, as we have seen in the previous chapter, provide evidence that perceptions and sensations are fed into parallel distributed processes that operate within and outside awareness (Kihlstrom, 1998 ; Kline, 1998). It has furthermore been established that neither the effective recognition and memorization of meaningful events, nor the traces of these processes in experimentally measured brain activity, do necessarily require awareness. These findings together with data showing that efficient sensory processing in the domain of vision is, as we have seen, possible in the absence of conscious experience, or that humans are capable of appropriate behavior with regard to objects they are not aware of, strongly suggests that the main functional role of subliminal perceptual processes may be that of preserving a maximum of incoming signals without overtaxing attentional capacities. Different levels of awareness may be mediated via neural signal exchanges at different levels of brain processing. These signal exchanges may be triggered by the sensory input itself and by other, related inputs that may or may not, depending on circumstances, be relevant to immediate behavior or ongoing awareness.

*All signals not relevant to ongoing awareness must be processed in parallel outside awareness*

It can be assumed that everything that is going on in the immediate environment of a person or an animal is of potential significance to ongoing representational processes. However, a great deal of decision making in every day life occurs without us becoming aware of what is going on. What is meant here is not only routine-like or automatic behavior in the classic sense, referring to learned skills that become automatized after learning (Kihlstrom, 1987). Knowledge outside awareness describes a far more complex domain than that of so-called automatic behavior, the « doing without thinking » that takes place once we have learnt to accomplish a specific task. In addition to that, we quite often make decisions or form judgements about objects, events, or circumstances although we cannot always articulate the way in which we have processed the seemingly relevant information, or even what that information was that has determined the judgements we make. Nonetheless, in many of these situations we find that our decisions or judgements are perfectly accurate and reliable. In fact, the ability of the human brain to generate representations in the absence of awareness, or subliminal knowledge if you wish, can explain why in so many situations in life we may quite safely rely on our intuition without having to make the effort of explicit reasoning. In a review article on this topic, Lewicki, Hill, & Czyzewska (1992) concluded that non-conscious information processing is not only faster, but also capable of generating multidimensional knowledge of interactive relations between variables that are too sophisticated to be processed inside awareness.

However, if we can merely agree on the ecological necessity that all incoming signals be processed by the brain, it becomes clear why all of this processing cannot take place in ongoing awareness. In fact, the reception, processing, and storage of information at subliminal levels would be the only efficient way of dealing with the potential relevance of a multitude of external signals and internal representations which do not need to be made available to immediate conscious processing. This inherent need for efficient integration of potentially but not necessarily immediately relevant information would therefore make it quite clear why we find subliminal perceptual phenomena at the detection (sensory), at the identification (recognition), and at the retrieval (memory) level. Kihlstrom (1987) suggested that awareness is basically dissociated from the different perceptive-cognitive functions such as discriminative responses to sensory inputs and other perceptual skills, memory, and even higher mental processes involved in judgements, decision making, or problem solving. On the basis of data relating to phenomena of subliminal perception and hypnotic alterations of consciousness, he introduced a taxonomy of what he referred to as « the cognitive unconscious » .

*Does the brain know more than we do ?*

Kihlstrom emphasized that humans seem to be able to perform cognitive analyses on information which is not itself accessible to awareness by means of automatized and unconscious procedural knowledge, and suggested a tripartite division of the « cognitive unconscious » into « truly unconscious », « preconscious », and « subconscious » parallel processes. These three would run in parallel with another, truly conscious, processing stream that generates declarative knowledge structures. Kihlstrom's theory thus suggests four parallel processes to account for the ways in which the brain generates knowledge at different levels of awareness. It thus implies that there are not two states, one where information is processed supraliminally (awareness), and one where the information that is being processed remains strictly subliminal (unawareness), but four levels where information is flowing through processing units which generate representations above or below some kind of threshold. Mechanisms that would explain how information passes these thresholds or, conversely, how it can be suppressed from a given level are not suggested in Kihlstrom's theory of the cognitive unconscious. It furthermore states that there is knowledge that will always remain unconscious in the strict sense of being « inaccessible to awareness under any circumstances ». By this statement, Kihlstrom assumes that the brain knows definitely more than we will ever be able to, and excludes the possiblity that the human brain might have the capacity to generate metarepresentations of subliminal knowledge at any time through context-sensitive adaptive learning. In principle though, such learning should be possible via neural mechanisms that bind distributed signals into coherent representations across levels of awareness, provided these signals match in the domain of knowledge and in the domain of time.

*Neural mechanisms to account for shifts in levels of awareness*

Neural network theory which regards the brain as a knowledge generating machine with multiple, parallel distributed unit structures is a valuable conceptual support to further our understanding of how different levels of representation may produce coherently organized knowledge structures where learning generates knowledge contents that are by nature subliminal but can, through further processing, be made available to ongoing awareness at any moment and at all processing levels. Neural networks have the capacity to generate metarepresentations of subliminal knowledge, and there seems to be no reason why the human brain could not do the same. In other words, there seems to be no good reason to assume that there is something like « truly unconscious », or « truly subliminal » representations in the strict sense of these representations being inaccessible to phenomenal awareness under any circumstances, as assumed in Kihlstrom's theory.

In formal neural networks, cells can become subliminally active if they receive priming signals that sensitize or modulate their actual respone or responsiveness by preparing them to react more quickly and vigorously to subsequent bottom-up inputs that match the priming signals (Carpenter & Grossberg, 1987 ; 1991). Perceptual knowledge of a visual enivronment, for example, would require that subliminal mechanisms be present in every cortical area wherein learning can occur, since without such mechanisms, any learned knowledge would be rapidly degraded and subject to what Grossberg refers to as «catastrophic forgetting » (Grossberg, 1999). Neural network models specifically developped to account for subliminal priming effects (Taylor, 1996) suggest modifications of neural reaction times to subsequent inputs, according to whether or not there are traces of subliminal processing of earlier input. Such models use parallel processing modules, or cell assemblies, with different lateral connectivity and output functions. Their functional properties are consistent with the hypothesis that the human brain uses parallel codes for the representation of contents or knowledge, and that these codes generate awareness when the discharges of functionally related neurons match in the domain of knowledge and in the domain of time. Such matches may be reflected by transient synchronization of the neural discharges (Singer, 1998).

Subliminal brain mechanisms may serve the purpose of boosting relevant, supra-liminal bottom-up signals and suppressing irrelevant signals at the appropriate time, and thus lead to a constant updating of current representational knowledge within and outside awareness. Recent data on temporal summation at dendrites of hippocampal neurons in the rat (Margulis & Tang, 1998), obtained with a technique where the strengths, sites, and timing of dendritic inputs can be controlled with precision, reveal that the temporal integration of synaptic inputs can readily switch between subthreshold an suprathreshold summation. The findings seem to suggest that active conductances in concert with passive cable properties in biological neural networks may serve to boost coincident synaptic inputs and attenuate or suppress noncoincident inputs. Such properties of synaptic transmission are consistent with the idea of brain mechanisms that generate interactions between context-specific, and temporally related, subliminal and supraliminal signals.

In the light of growing evidence in this domain, and the possibilities offered by neural network theory, it seems that a clear path towards a mechanistic working hypothesis, or model, of subliminal perceptual processes can now be traced. The working hypothesis introduced here revises some of the key statements formulated in Kihlstrom's (1987) theory of « the cognitive unconscious » by stating that subliminal perceptual input is processed and represented in all areas of the brain that can generate learning, that these subliminal representations can be made available to awareness under specific circumstances by way of adaptive neural mechanisms, and that supraliminal representations can be suppressed from awareness under other circumstances by similar mechanisms.

*Subliminal representation*

The question here is how to account for the way in which subliminal input traces can be processed and stored in neural structures without interfering with ongoing supraliminal processes or, more importantly, without destroying or changing representations that are already stored in these structures. Grossberg (1999) proposed a synthesis of how Adaptive Resonance Theory (ART) can be used to build cortical models that are able to account for subliminal learning and knowledge representation in the brain by satisfying a limited set of functional properties which strike by their ingeniosity as well as by their simplicty. The neural network property that can be used to explain the processing and storage of perceptual input outside phenomenal awareness is referred to as Bottom-Up Automatic Activation by Carpenter & Grossberg (1991) as well as by Grossberg (1999). Bottom-Up Automatic Activation describes a mechanism where a group of cells within a given structure becomes supraliminally active when it receives bottom-up signals that may themselves be subliminal, i.e. not consciously perceived. These signals are then multiplied by adaptive weights that represent long-term memory traces and influence the activation of the cells at a higher processing level. Grossberg (1999) originally proposed Bottom-Up Automatic Activation to account for the way in which pre-attentive processes generate learning in the absence of top-down attention or expectation. It appears that this mechanism is equally well suited to explain how subliminal signals may trigger supraliminal neural actvities in the absence of phenomenal awareness. In fact, if we consider cases where Bottom-Up Activation generates supraliminal brain signals or representations with adaptive weights near or at zero, we would have a good candidate mechanism to explain how the brain manages to process perceptual input that is either not relevant to ongoing awareness, or cannot be made available to further processing (i.e. awareness) because of some kind of lesion. Such a mechanism would provide a good model structure for blindsight phenomena in patients with brain lesions and in normal subjects. These phenomena are briefly reviewed in the first chapter of this article, describing observations where subjects are perfectly able to make correct judgements about specific characteristics of visual stimuli but are not able to tell what they see.

*Breakthrough to awareness*

Subliminal perceptual input could in principle be made available to awareness whenever a subliminal bottom-up signal or representation is multiplied by a number of adaptive weights that is sufficient to activate top-down expectation signals or representations that match the bottom-up representation in the domain of time and in the domain of contents. Any positive match would then confirm and amplify this bottom-up representation supported by a sufficient number of learned long-term memory traces (Grossberg, 1999) and thereby trigger a selective, and synchronized process where initially subliminal representations are incoporated in ongoing awareness. This matching rule was originally used in neural networks based on ART to solve what Grossberg refers to as « the attention-preattention interface problem » (Grossberg, 1999) by enabling preattentive bottom-up processes to use some of the same circuitry that attentive processes use, even before these latter may come into play, in order to stabilize cortical development and learning. The top-down matching rule generates, in a very general sense, feed-back resonances between bottom-up and top-down signal exchanges which can rapidly bind information at multiple levels of brain processing into context-sensitive representations of objects and events. It provides a good model structure for subliminal semantic priming (e.g. Marcel, 1983), reviewed in the first chapter of this article. The top-down matching hypothesis is also consistent with psychophysical data showing that initially subliminal visual targets can be made available to awareness, or « break through » to supraliminal levels of processing, when they are embedded in the appropriate perceptual object context (e.g. Dresp & Grossberg, 1997 ; 1999). This « breakthrough » of subliminal visual processing requires a certain amount of learning, or practice (Dresp, 1998), which is consistent with the ART network prediction that bottom-up representations are progressively enhanced as resonance takes hold (Grossberg, 1999).

*Suppression from awareness*

Whenever a bottom-up representation is either not multiplied by a sufficient number of learned long-term memory traces, or does not match a top-down representation, or expectation signal, activated at a further level of processing, the bottom-up representation would either remain outside awareness, or be suppressed from awareness. This is a logical consequence from the resonance principles and matching rules used in ART neural networks (Grossberg, 1999). In the domain of the subliminal perceptual phenomena reviewed in chapter one of this article, this would refer to those cases where awareness is found to have a negative, or suppressive, influence on the processing of bottom-up signals, as found in certain cases of psychodynamic suppression effects where awareness makes it harder to process affect- or drive-related stimuli efficiently (Bornstein, 1990). A recent study by Wolfe & Alvarez (1999) has shown that visual search is more efficient, or faster, when the observers are unaware of what to attend to. In all of these cases, it seems as if guided awareness may in some cases create rigid top-down expectations that will interfere with letting a positive resonance between bottom-up and top-down signal exchanges take hold naturally.

Negative resonance between bottom-up and top-down signal exchanges in massively parallel neural network architectures can provide an ideally suited model structure for one of the greatest enigma in research on subliminal psychodynamic activation effects, or hypnotic alterations of consciousness : the hypnotic control of physical pain (e.g. Chaves & Dworkin, 1997). In fact, within the framework of ART network structures it becomes relatively straightforward to see how hypnotic states, where overall awareness may be shifted away from the general context and environment and focussed on the symbiotic fantasies induced by the verbal suggestions of an expert clinician, activates powerful top-down representations that can almost totally suppress somatosensory bottom-up signals from ongoing awareness through progressively enhanced, negative resonance. Thus, what was hitherto considered as mere humbug by many could now, in principle, be confronted with the functional predictions of a cortical neural network in a general scientific framework of cognitive neuroscience and clinical research.

Kihlstrom's (1987) claim that perceptions and representations embedded in immediate, ongoing awareness can be fed into parallel distributed processes that operate outside awareness was couched within the classic information-processing conception of human cognition, and it was basically correct. However, the functional principles and concepts from current neural network theory that are discussed herein invite us to have a new look at Kihlstrom's theory of the « cognitive unconscious », and to revise some of its statements in the following conclusion.

**Conclusions**

Kihlstrom (1987) suggested that awareness is basically dissociated from the different perceptive-cognitive functions such as discriminative responses to sensory inputs, or memory processes. This claim is supported by some of the empirical evidence reviewed in the first chapter of this article. Data on human blindsight phenomena, for example, reveal a dissociation between successful sensory discrimination and verbal commentaries which acknowledge no awareness of the discriminanda (Weiskrantz, 1995), and thereby tend to confirm Kihlstrom's claim. Furthermore, perceptual learning and representational processing during early childhood is generated at a fast rate quite a long time before awareness and intentionality, or goal-directedness of behavior begin to emerge (Piaget, 1967 ; Flavell, 1999). However, despite some agreement that awareness may not be specific for a given perceptual or cognitive function, the idea of a functional dissociation in the strict sense seems to be hard to reconcile with some recent data. For example, research on object perception suggests that a specific form of awareness, mediated through visual attention, is required for visual object identification (Treisman, 1998), or that visual awareness and object identification are associated with activities in the same functionally identified cortical area (Bar & Biederman, 1999). Furthermore, interactions between the identification of words and guided visual attention have been found (Stone, Ladd, Vaidya, & Gabrieli, 1998). Experimental data on classical conditioning reveal that simple delay conditioning of a reflex response occurs without awareness whereas trace conditioning, a variant of the standard paradigm where the temporal interval is given between the offset of the conditional stimulus and the onset of the unconditional stimulus, of the same response does require awareness. An interesting case of selective perceptual awareness in a patient with right parietal brain damage which is difficult to reconcile with a strict dissociation of awareness from perceptual or cognitive skills has been reported recently (Chatterjee & Thompson, 1998). While motorically aware and actively engaged in lifting weights with her left hand, the patient was unaware of incremental changes in the weights. Another case of fragmented bodily awareness describes a patient with right frontal lesion who ocasionally perceives a « ghost » hand which copies the previous position of his left hand with a temporal lag, but follows the movement patterns of his right hand (Hari et al., 1998). These findings may be interpreted as some kind of support for the idea that a specific form of awareness appears to be associated with a specific perceptual or cognitive function. With regard to Kihlstrom's claim of functional dissociation, the working hypothesis adopted here in this article would be that awareness is generated in all cortical areas that may generate learning and that it is therefore conceivable that some of the interactions between representations generated outside awareness and knowledge inside awareness are, indeed, specific for a given cortical area or learning domain.

The statement that there is knowledge that will always remain unconscious in the strict sense of being « inaccessible to awareness under any circumstances » (Kihlstrom, 1987) implies that the brain knows more than we do in the sense that some part of the knowledge generated by the brain is once and for always excluded from conscious representation and communication via metaphor. As a consequence, it excludes the possiblity that the human brain might have the capacity to generate metarepresentations of subliminal knowledge at any time through context-sensitive adaptive learning. It is difficult to reconcile this idea with experimental findings showing that initially subliminal visual targets can be made available to awareness, or « break through » to supraliminal levels of processing, when they are embedded in the appropriate perceptual object context (Dresp & Bonnet, 1995 ; Kapadia, Ito, Gilbert & Westheimer, 1995 ; Dresp & Grossberg, 1997 ; 1999), and with some of the functional properties of neural network architectures designed to model how the human brain generates perception, representation, awareness, and learning (Grossberg, 1999). The working hypothesis presented herein is based on Grossberg's latest developments on Adaptive Resonance Theory (ART) and suggests that signals or knowledge processed outside awareness can be made available to awareness via positive resonance between bottom-up and top-down signal exchanges in massively parallel neural network architectures, povided these signals match in the domain of time and in the domain of knowledge. Such resonance generates adaptive learning by selectively synchronizing and binding distributed information at multiple levels of brain processing into context-sensitive representations of knowledge and events.

Finally, the suggested tripartite division of the « cognitive unconscious » into « truly unconscious », « preconscious », and « subconscious » parallel processes (Kihlstrom, 1987) seems to have lost the heuristic value it may have had twelve years ago. In view of recent data and theory, it appears more fruitful to consider that representations may be generated inside and outside awareness in all cortical areas that generate learning, and that adaptive mechanisms that produce interactions between actual context and stored knowledge traces determine what is made available to awareness and what is suppressed from awareness at a given time. The neural network approach used herein to develop the working hypothesis for subliminal perceptual processes was originally developped to account for mechanisms of attention, cortical development, and learning (Carpenter & Grossberg, 1991). It focusses on a combination of intracortical and intercortical pathways. The hypotheses formulated therein are supposed to be sufficiently backed up by neurophysiological data (see Grossberg, 1999, for references), which makes the approach biologically plausible. This target article provides, among other things, an attempt to illustrate the extent to which such a neural network approach may be adapted to handle phenomena at the interface between subliminal and supraliminal representational processing, and to generate explanations and models that will lend support to further experimental research in this domain, which is likely to teach us a great deal about the brain beyond the « supraliminal approach » to perceptual and cognitive function.